\documentclass[aps,prl,twocolumn,showpacs]{revtex4}
\usepackage{amssymb}
\usepackage{graphicx}
\usepackage{amsmath}
\usepackage{amsfonts}
\begin{document}

\title{Proximity effect and Anomalous metal state in a model of mixed metal-superconductor grains. 
 }
\date{\today}
\author{Tai-Kai Ng}
\affiliation{Department of Physics, Hong Kong University of Science and Technology, Clear Water Bay, Hong Kong, China}

\begin{abstract}
       Motivated by the suggestions that the coexistence of superconducting and metallic components is crucial to the formation of the anomalous metal state in thin film systems, we study in this paper a model of mixed metallic and superconducting grains coupled by electron tunneling - the metallic grains are expected to become superconducting because of proximity effect in a mean-field treatment of the model. When quantum fluctuations in relative phases between different grains are taken into account, we show that the proximity effect can be destroyed and the metallic and superconducting grains become "insulating" with respect to each other when the charging energy between grains are strong enough and tunneling between grains are weak enough, in analogy to superconductor-insulator transition in pure superconducting grains or metal-insulator transition in pure metallic grains. Based on this observation, a physical picture of how the anomalous metal state may form is proposed. An experimental setup to test our proposed physical picture is suggested. 

\end{abstract}

\pacs{74.45.+c, 74.78.-w, 74.81.Bd}
\maketitle

\subsubsection{introduction}
   The discovery of anomalous metal states in a variety of two dimensional electronic systems poses a strong challenge to the condensed matter physics community\cite{rmp}. The anomalous metal state can be found in a wide variety of materials including conventional BCS superconductors, cuprate thin films\cite{rmp} and more recently, layered superconductors\cite{topo} and 3D system\cite{3d} by tuning some (non-thermal) parameters which normally drive the system from a superconducting state towards insulator. At an intermediate regime, an anomalous metal can be observed where the system remains metallic down to lowest observable temperature with conductance that can be order of magnitude above or below the quantum conductance $e^2/h$\cite{rmp}. The state has been observed in non-uniform, granular systems as well as homogeneous, disordered films\cite{rmp,topo}.   

   Motivated by these discoveries and suggestions that coexistence of metallic and superconducting components is crucial to the formation of anomalous metal state\cite{rmp,fl,sz,sk,pnas,art1, art2,art3}, we consider in this paper a model of mixed metallic and superconducting grains coupled by electron tunneling, and investigate whether the proximity effect\cite{proximity} which induces superconductivity on the metallic grains can be destroyed by quantum fluctuations, leading to zero temperature metallic behaviour. We consider systems where disorder leads to separate weakly coupled regions with size $\geq \xi_0$ with attractive interaction existing in some regions, and repulsive interaction in others, where $\xi_0$ is the superconductor coherence length. In this case, the system can be modelled as an effective mixture of metallic and superconducting grains coupled by electron tunneling. 
   
   \subsubsection{Model}
   We consider conventional BCS superconductor and metallic grains in this paper. Following Refs.\cite{granular1, granular2}, the imaginary time Lagrangian describing such a system is (we set $\hbar=e=1$ in the following discussions) $L=\sum_i L_{si}+\sum_j L_{mj} +\sum_{i,j}(T_{ij}+C_{ij})$, where
   \begin{subequations}
   \begin{equation}
    \label{L2}
    L_{mi}=\sum_{k\sigma}c^+_{ik\sigma}(\tau)
    \left(\frac{\partial}{\partial\tau}+\xi_k-iV_i(\tau)\right)
    c_{ik\sigma}(\tau)
   \end{equation}
   is the Lagrangian describing metallic grain $i$. $c_{ik\sigma}(c^+_{ik\sigma})$'s are spin-$\sigma$ electron operators operating on electronic eigenstate $k$ in grain $i$. The electrons are also coupled to a $\tau$-dependent scalar field $V_i(\tau)$ which we shall see describes charging effects\cite{granular1}.
   \begin{eqnarray}
   \label{L1}
       L_{si} & = & L_{mi}+ U\lambda^*_i(\tau)\lambda_i(\tau)  \\ \nonumber
       & & +\left[i\Delta_i(\tau)\left(\lambda^*_i(\tau)-\sum_k c^+_{ik\uparrow}(\tau)c^+_{i-k\downarrow}(\tau)\right)+c.c.\right]
   \end{eqnarray}
   is the Lagrangian for superconducting grain $i$ where $\lambda_i(\lambda^*_i)$ are the Cooper pairing fields with the constraint $\lambda^*_i(\tau)\equiv\sum_k c^+_{ik\uparrow}(\tau)c^+_{i-k\downarrow}(\tau)$ imposed by the Lagrange multiplier field $\Delta_i(\Delta^*_i)$. We assume that the Cooper pairing field $\lambda_i(\tau)$ is uniform within a grain and fluctuates only in time $\tau$. This is justifiable if the grain sizes $L$ are comparable with the superconductor coherence length $\xi_0$. 
   \begin{equation}
    \label{L3}
    C_{ij}=\frac{1}{2c_{ij}}\left(V_i(\tau)-V_j(\tau)\right)^2   
   \end{equation}
   describes the charging effect between grains $i$ and $j$ where $c_{ij}$ is the capacitance.
   \begin{equation}
    \label{L4}
    T_{ij}=\sum_{kp\sigma}(t^{(ij)}_{kp}c_{ik\sigma}^+(\tau)c_{jp\sigma}(\tau)+c.c.)
   \end{equation}
   describes single electron tunneling between state $k$ in grain $i$ and state $p$ in grain $j$ with $t^{(ij)}_{kp\sigma}$ being the corresponding tunneling matrix element. The matrix elements $t^{(ij)}_{kp}=(t^{ji}_{pk})^*$ are random and uncorrelated because of random configuration and shape of grains and satisfy 
   \begin{equation}
    \label{L5}
    \langle t^{(ij)}_{kp}t^{(i'j')}_{k'p'}\rangle\sim \frac{t^2}{V} \delta_{ii'}\delta_{jj'}\delta_{kk'}\delta_{pp'}.
   \end{equation}
   where $V=$ volume of grain. We shall assume $t$ to be a real number in our discussions.
   \end{subequations}
  
   We shall consider only phase fluctuations in this paper and write $\lambda_i(\tau)=|\lambda_i|e^{i\phi_i(\tau)}$. Performing a gauge transformations $c_{ik\sigma}(\tau)\rightarrow c_{ik\sigma}(\tau)e^{i\frac{\phi_i(\tau)}{2}}$ and $\Delta_i(\tau)\rightarrow\Delta_i(\tau)e^{-i\phi_i(\tau)}$, we obtain
   $\lambda_i(\tau)\rightarrow|\lambda_i|$, $\Delta_i(\tau)\rightarrow|\Delta_i|$,
   \begin{subequations}
   \begin{equation}
    V_i(\tau)\rightarrow V_i(\tau)-\frac{1}{2}\frac{\partial}{\partial\tau}\phi_i(\tau)
   \end{equation}
    in $L_{mi}$ and $L_{si}$ and 
   \begin{equation}
    T_{ij}\rightarrow\sum_{kp\sigma}(t^{(ij)}_{kp}e^{\frac{i}{2}(\phi_i(\tau)-\phi_j(\tau))}c_{ik\sigma}^+(\tau)c_{jp\sigma}(\tau)+c.c.).
   \end{equation}
   \end{subequations}
   Applying the Josephson relation $V_i(\tau)=\frac{1}{2}\frac{\partial}{\partial\tau}\phi_i(\tau)$ and integrating out the fermion fields, we obtain an effective action for $\{\phi_i\}$ fields\cite{granular1} describing dynamics of the tunneling junction between different grains,
   \begin{eqnarray}
   \label{action1}
    S(\{\phi_i\}) & = & \sum_{i,j}\frac{1}{2E_{ij}}\int_0^\beta d\tau\left(\frac{\partial}{\partial\tau}(\Delta\phi_{ij}(\tau)\right)^2
      \\ \nonumber
      & & +Trln(M(\{\phi_i\})),
   \end{eqnarray}
   where $E_{ij}=4/c_{ij}$ is the charging energy, $\Delta\phi_{ij}(\tau)=\phi_i(\tau)-\phi_j(\tau)$ and $M(\{\phi_i\})=M_0+T(\{\phi_i\})$. $M_{0kp}^{(ij)}=\delta_{ij}\delta_{kp}m_{0ik}$, with
   \begin{subequations}
   \label{m}
   \begin{equation}
    \label{m0}
    m_{0ik}=\begin{pmatrix}
    \frac{\partial}{\partial\tau}+\xi_{ik} & |\Delta_i| \\
   |\Delta_i|  & -\left(\frac{\partial}{\partial\tau}-\xi_{i-k}\right)
   \end{pmatrix}
   \end{equation}
   and
   \begin{equation}
    \label{T}
    T_{kp}^{(ij)}(\{\phi_i\})=\begin{pmatrix}
    t_{kp}^{(ij)}e^{\frac{i}{2}\Delta\phi_{ij}(\tau)} & 0 \\
   0  & -t^{(ji)}_{pk}e^{-\frac{i}{2}\Delta\phi_{ij}(\tau)}
   \end{pmatrix}.
   \end{equation}
   \end{subequations}
   In previous works\cite{granular1, granular2, CK, NL} the matrix $Trln(M(\{\phi_i\})$ is expanded in a power series of $T(\{\phi_i\})$, and only the second order term ($=\frac{1}{2}G_0T(\{\phi_i\})G_0T(\{\phi_i\})$) is kept, resulting in an effective Lagrangian to second order in $e^{\frac{i}{2}\Delta\phi_{ij}(\tau)}$\cite{granular1,granular2}, where $G_0=M_0^{-1}$. 
   We shall follow this treatment in our paper, except replacing $G_0$ by $G$, where $G=(M_0+T_0)^{-1}$ is the electron Green's function taking into account the electron tunneling effects in the absence of phase fluctuations, where $T_0$ is given by Eq.\ (\ref{T}) with $\Delta\phi_{ij}(\tau)=0$ $\forall i,j$, corresponding to a mean-field treatment of proximity effect where phase fluctuations are neglected.
   
   \subsubsection{Renormalized Green's functions}
   
   The computation of $G$ is non-trivial because of the random matrix elements $t_{kp}^{(ij)}$. We shall employ a self-consistent Born approximation in the following, where we approximate 
   \begin{equation}
    \label{Born}
    G^{-1}=G_0^{-1}-\Sigma,  \,\  \Sigma\sim Tr(T_0GT_0).
   \end{equation}
   Using Eqs.(\ref{L5}) and (\ref{m}), we obtain $\Sigma^{(ij)}_{kp}=\delta_{ij}\delta_{kp}\Sigma_{ik}$, where
   \begin{equation}
    \label{se}
    \Sigma_{ik}(i\omega)=|t|^2\sum_{j=i+\delta}\frac{1}{V}\sum_p G_{jp}(i\omega).
   \end{equation}
    We have assumed nearest-neighbor tunneling only in writing down Eq.\ (\ref{se}) where $\delta=$ nearest neighbor sites. Notice that $G_{kp}^{(ij)}=\delta_{ij}\delta_{kp}G_{ik}$ in self-consistent Born approximation and $\Sigma_{ik}$ is independent of $k$. 
   
   To proceed further we consider identical metal- and superconductor- grains occupying the A- and B- sub-lattice sites of a regular (2D) square lattice. For nearest neighbor coupling there exists coupling only between superconductor and metallic grains. Assuming translational invariant solutions for $G$ in each sub-lattice, we obtain site-independent self-energies $\Sigma_{M(S)}(i\omega)$ for metal(superconductor) grains, respectively with corresponding Green's functions $G_{M(S)}(i\omega)=\frac{1}{V}\sum_k G_{M(S)k}(i\omega)$. Writing
   \begin{equation}
    \label{gg}
    G_{M(S)}(i\omega) = \begin{pmatrix}
    g_{M(S)}(i\omega) & -f_{M(S)}(i\omega)  \\
    -f_{M(S)}(i\omega) & g_{M(S)}(i\omega)
    \end{pmatrix},
   \end{equation}
   and defining $\Sigma_{M(S)}(i\omega)=4|t|^2g_{S(M)}(i\omega)$ and $D_{M(S)}(i\omega)=4|t|^2f_{S(M)}(i\omega)$,
   we obtain the self-consistent equations, 
   \begin{eqnarray}
    \label{scg}
    G_M(i\omega) & = & \frac{-i\pi N_M(0)}{\sqrt{(i\omega-\Sigma_M(i\omega))^2-|D_M(i\omega)|^2}}\times  \\ \nonumber
    & & \begin{pmatrix}
    i\omega-\Sigma_M(i\omega) & -D_M(i\omega)  \\
    -D_M(i\omega) & i\omega-\Sigma_M(i\omega)
    \end{pmatrix},  \\ \nonumber
     G_S(i\omega) & = & \frac{-i\pi N_S(0)}{\sqrt{(i\omega-\Sigma_S(i\omega))^2-|\Delta_0+D_S(i\omega)|^2}}\times  \\ \nonumber 
     & & \begin{pmatrix}
    i\omega-\Sigma_S(i\omega) & -(\Delta_0+D_S(i\omega))  \\
    -(\Delta_0+D_S(i\omega)) & i\omega-\Sigma_S(i\omega)
    \end{pmatrix}, 
   \end{eqnarray}
    $\Delta_0$ is the superconductor gap of the (unperturbed) superconducting grains and $N_{M(S)}(0)$ is the density of states of the metallic (superconducting) grains on Fermi surface. We have approximated $\frac{1}{V}\sum_k\rightarrow N(0)\int_{-D}^Dd\xi$ and take the limit $D\rightarrow\infty$ in deriving the above self-consistent equations. The details of the calculation can be found in the supplementary materials.
   
   The solution of the above self-consistent equations with $i\omega\rightarrow\omega$ can be written in the form
   \begin{equation}
    \label{scg1}
    G_{M(S)}(\omega) = \frac{-i\pi N_{M(S)}(0)}{\sqrt{\omega^2-|\Delta_{M(S)}(\omega)|^2}}\begin{pmatrix}
    \omega  & -\Delta_{M(S)}(\omega)  \\
    -\Delta_{M(S)}(\omega) & \omega
    \end{pmatrix},  
   \end{equation}
  where $\Delta_{M(S)}(\omega)$ is a $\omega$-dependent gap function. We first consider $\Gamma_M,\Gamma_S>>\Delta_0>\omega$, where $\Gamma_{M(S)}=4\pi|t|^2N_{M(S)}(0)$ are tunneling widths. In this limit, it is straightforward to show that $\Delta_M(\omega)\rightarrow\Delta_S(\omega)\sim\Delta_0$ and the system become a homogeneous superconductor with more-or-less uniform superconductor gap because of strong proximity effect. 
  
  In the other limit $\Gamma_M,\Gamma_S<<\Delta_0 (>>\omega)$, we obtain
  \begin{eqnarray}
   \label{Ds}
   \Delta_M(\omega) & = & \frac{\Gamma_S\Delta_S(\omega)}{\sqrt{|\Delta_S(\omega)|^2-\omega^2}+\Gamma_S}\sim\Gamma_S  \\ \nonumber
   \Delta_S(\omega) & \sim & 
   \frac{\Delta_0\omega}{\omega+i\Gamma_M}, (\omega>>\Delta_M(\omega))  \\  \nonumber
   & \sim & \frac{\Delta_0\Gamma_s}{\Gamma_S+\Gamma_M}, (\omega<<\Delta_M(\omega)).
  \end{eqnarray}
  Notice that the induced superconducting gap on the metallic grains $\sim \Gamma_S$ is much smaller than $\Delta_0$ in this limit and the system becomes highly in-homogeneous.
  
  We note that a parallel analysis on pure superconducting or pure metallic grains indicate that the Green's functions are {\em not} renormalized by electron tunneling in these cases. The details of these calculations are given in the supplementary materials.
  
  We now examine the properties of the Green's functions in the regime  $\Gamma_M,\Gamma_S<<\Delta_0$ in more detail. We first consider $G_S$ and $G_M$ at energy ranges $\Delta_M(\omega)<<\omega<<\Delta_0$. Using Eqs.\ (\ref{scg}) and\ (\ref{Ds}) we obtain 
     \begin{subequations}
     \label{gf0}
     \begin{eqnarray}
    \label{gf1}
    G_M(\omega) & \sim & -i\pi N_M(0)\begin{pmatrix}
    1  & 0  \\
    0 & 1
    \end{pmatrix},  \\ \nonumber
     G_S(\omega) & \sim & \frac{-i\pi N_S(0)}{\sqrt{(\omega+i\Gamma_M)^2-|\Delta_0|^2}}\begin{pmatrix}
    \omega+i\Gamma_M & -\Delta_0  \\
    -\Delta_0 & \omega+i\Gamma_M
    \end{pmatrix}, \\ \nonumber
    & \sim & G_1(\omega)+G_2(\omega)
    \end{eqnarray}
    where
    \begin{eqnarray}
     \label{gf2}
    G_1(\omega) & = & \frac{-i\pi N_S(0)}{\sqrt{\omega^2-|\Delta_0|^2}}\begin{pmatrix}
    \omega & -\Delta_0 \\
    -\Delta_0 & \omega
    \end{pmatrix} \\ \nonumber
    G_2(\omega) & = & -\frac{i\pi N_S(0)\Gamma_M}{|\Delta_0|}\begin{pmatrix}
    1 & 0 \\
    0 & 1
    \end{pmatrix}. 
   \end{eqnarray}
   We find that $G_S$ behaves as a superconductor with gap $\Delta_0$ except that the quasi-particles acquire a finite life time $\sim\Gamma_M$. The finite life time effect leads to the additional metallic-like component $G_2$.
   
   These results are direct consequences of the large difference between $\Delta_M$ and $\Delta_S$. In the energy range $\omega>\Delta_M$ the proximity effect is not yet effective and the metallic grains behave as metals. Electrons in the superconducting grains with energy $\Delta_S>\omega>\Delta_M$ can tunnel into the metallic grains resulting in a finite lifetime $\sim\Gamma_M$ and the appearance of $G_2$ in this energy range. The system becomes a fully-gaped superconductor only at energy range $\omega<\Delta_M$ where we obtain
   \begin{eqnarray}
    \label{gf3}
    G_{M(S)}(\omega) & \rightarrow & \frac{-i\pi N_{M(S)}(0)}{\sqrt{(\omega)^2-|\Delta'_{M(S)}|^2}}\begin{pmatrix}
    \omega  & -\Delta'_{M(S)}  \\
    -\Delta'_{M(S)} & \omega
    \end{pmatrix}, 
   \end{eqnarray}
     with $\Delta'_M\sim\Gamma_S$ and $\Delta'_S\sim\frac{\Delta_0\Gamma_s}{\Gamma_S+\Gamma_M}$. 
     \end{subequations}

\subsubsection{Effective phase model}
   Following previous works\cite{granular1, granular2}, we expand the logarithmic term in phase action $S(\{\phi_i\})$ (Eq.\ (\ref{action1})) to second order, with the bare Green's functions $G_0$ replaced by the renormalized Green's function $G$, leading to the well studied phase action
   \begin{subequations}
   \label{action}
   \begin{eqnarray}
   \label{action3}
    S(\{\phi_i\}) & \rightarrow & \sum_{i,j=i+\delta}\int_0^\beta d\tau\left\{\frac{1}{2E_{ij}}\left(\frac{\partial}{\partial\tau}(\Delta\phi_{ij}(\tau))\right)^2\right.
      \\ \nonumber
     & & +\int_0^{\beta}d\tau'\alpha_{ij}(\tau-\tau')\cos(\frac{\Delta\phi_{ij}(\tau)-\Delta\phi_{ij}(\tau')}{2})  \\  \nonumber
      & & \left.-J_{ij}\cos\left(\Delta\phi_{ij}(\tau)\right)\right\}.
   \end{eqnarray}
    The two terms $\alpha_{ij}$ and $J_{ij}$ describes normal (single) electron tunneling and Josephson coupling between grains $i$ and $j$, respectively. For our regular lattice model with metallic and superconducting grains occupying different sub-lattices, $E_{ij}\rightarrow E_0$ where $E_0$ is the charging energy between the superconductor and metallic grains,
     \begin{eqnarray}
      \label{ab}
      \alpha_{ij}(\tau) & = & 2|t|^2g_i(\tau)g_j(-\tau)  \\ \nonumber
      & \rightarrow & \alpha_{MS}(\tau)=2|t|^2g_M(\tau)g_S(-\tau),  \\ \nonumber
      J_{ij} & = & 2|t|^2\int^{\beta}_0d\tau f_i(\tau)f_j(-\tau)  \\ \nonumber
       & \rightarrow & J_{MS}=2|t|^2\int^{\beta}_0d\tau f_S(\tau)f_M(-\tau),
     \end{eqnarray}
     \end{subequations}
     $g_i(\tau)$ and  $f_i(\tau)$ are the normal and anomalous Green's function on site $i$ at imaginary time. Notice that $\alpha_{ij}\rightarrow\alpha_{MS}$ and $J_{ij}\rightarrow J_{MS}$  as $i,j$ always connect between superconductor and metallic grains in our regular lattice model.
     
     Using Eqs.\ (\ref{gf0}),  we obtain for energy range $\Delta_M(\omega)\sim\Gamma_S<<\omega<<\Delta_0$,
  \begin{eqnarray}
      \label{j&a}
      J_{MS} & \sim & \frac{h}{2 e^2R_{SM}}\int^{\beta}_0d\tau\Delta_0\Gamma_SK_0(\Delta_0\tau)K_0(\Gamma_S\tau)   \\ \nonumber
      \alpha_{MS}(\tau) & \sim & \frac{h}{2e^2R_{SM}}\left(\Delta_0K_1(\Delta_0\tau)
      +\frac{\Gamma_M}{\Delta_0}\frac{\pi kT}{\sin(\pi kT\tau)}\right)  \\ \nonumber
      & & \times\frac{\pi kT}{\sin(\pi kT\tau)}\sim\alpha_{MS}(\frac{kT}{\sin(\pi kT\tau)})^2
  \end{eqnarray}
  where $R_{ab}^{-1}=4\pi e^2|t|^2N_a(0)N_{b}(0)$ is the normal state tunneling conductance between grains $a$ and $b$. We have put back the Planck's constant $\hbar$ and electric charge $e$ into the expressions for $J$ and $\alpha(\tau)$.
  The first term in $\alpha_{MS}(\tau)$ comes from $G_1$ and is much smaller than the second term for $\Delta_0\tau>>1$.
 
  At energy range $\omega<<\Gamma_S$, $G_2$ is absent and the Action has the same form as\ {(\ref{action3}), except that the single electron tunneling term is absent and $J_R$ is further renormalized as $\Delta_0\rightarrow\Delta'_S$. 
  
   We note that the action\ (\ref{action}) with $J_{MS}\rightarrow J_{SS}$ and $\alpha_{MS}\rightarrow \alpha_{MM}$ where
   \begin{eqnarray}
      \label{rsj}
  J_{SS} & = & \frac{h}{2e^2R_{SS}}\int^{\beta}_0d\tau\left[\Delta_0K_0(\Delta_0\tau)\right]^2  \\ \nonumber
    \alpha_{MM} & = & \frac{h}{2e^2R_{MM}}
  \end{eqnarray}
   has been used to study superconductor-insulator transition between identical resistance-shunted superconductor junctions (RSJ)\cite{CK,NL}. It was observed that at zero temperature 
   the Josephson coupling between grains $J_{SS}$ is renormalized to zero and superconductivity is destroyed when both $J_{SS}/E_0$ and $\alpha_{MM}$ are small enough\cite{CK}. These results are summarized briefly in the supplementary materials. We expect that the zero-temperature phase diagram of our mixed grain model shared similar behaviour except that in the RSJ model, the local superconducting order parameter $\Delta_i$'s remain intact as $J_{SS}\rightarrow0$, but $\Delta_M\rightarrow0$ when $J_{MS}\rightarrow0$ in our model, as the proximity effect is destroyed when the Josephson coupling $J_{MS}$ vanishes\cite{t00}. Notice that the regime $\omega<\Delta_M$ is unimportant as far as the $J_{MS}\rightarrow0$ transition is concerned as the regime vanishes when $\Delta_M\rightarrow0$.
   
   Performing the integrals for $J_{MS}$ and $J_{SS}$ we find that
   \begin{eqnarray}
   \label{ratio}
   J_{MS} & \sim & (\frac{\Gamma_S}{\Delta_0})(\frac{N_{M}(0)}{N_{S}(0)})\ln(\frac{\Delta_0} {\Gamma_S})J_{SS}<<J_{SS}  \\ \nonumber
   \alpha_{MS} & \sim & (\frac{\Gamma_M}{\Delta_0})(\frac{N_{S}(0)}{N_{M}(0)})\alpha_{MM}<<\alpha_{MM}
   \end{eqnarray}
   in the regime $\Gamma_S\sim\Gamma_M<<\Delta_0$, suggesting that the non-superconducting regime is much expanded in our model compared with the RSJ model. As illustration a schematic zero temperature phase diagram of our model is shown in Fig. (1), assuming that the mixed grain model has the same $\Delta_0$ and same $\Gamma_S=\Gamma_M=\Gamma$ as the corresponding RSJ model with $R_{SS}=R_{MM}$, following previous results for the RSJ model\cite{CK,NL}. 

    \begin{figure} 
\centering
\includegraphics[scale=0.5]{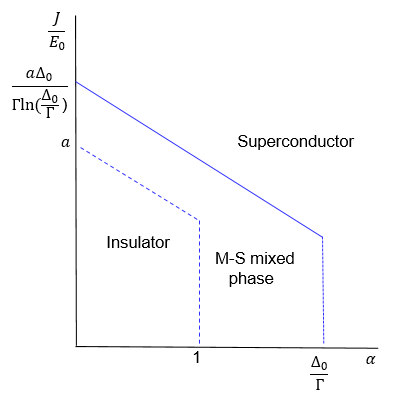}
\caption{schematic zero temperature phase diagram of the mixed grain model. $a$ is a number of order O(1). The dash (solid) line represents the transition in the RSJ (mix grain) model with $\alpha$ and $J$ given by Eq.\ (\ref{rsj}).} 
\end{figure}

  \subsubsection{Anomalous metal state}
     The coexistence of metallic and superconducting components in granular superconductors has been proposed as a necessary ingredient describing the anomalous metal state\cite{fl,sz,sk,pnas}. We shall discuss how this may happen in our model.

     In realistic granular films the attractive / repulsive interaction regions are randomly distributed and the effective metallic/superconducting grains are connected randomly. We shall make use of the results we obtained in previous sections to propose a qualitative picture for the behaviour of this random grain system, assuming that the superconducting grains have the same gap magnitude $\Delta_0$ and the grains all have large enough charging energy $E_{0}$. We shall also assume that the tunneling widths between grains is random with same mean value $\Gamma$ and standard deviation $\sigma_{\Gamma}<\Gamma$ for both metallic and superconducting grains and consider the behaviour of the system as a function of $\Gamma$. 
     
     We expect:
    
    (1) When $\Gamma\geq\Delta_0$ the system behaves as a macroscopic superconductor with more-or-less uniform superconducting gaps.
    
    (2) When $\Gamma< \Delta_0$ non-uniformity develops in the superconducting gaps $\Delta_i$ in the system with the metallic grains having smaller gaps than superconductor grains.
    
    (3) When $\Gamma$ decreases further proximity effect is destroyed in some metallic grains and these grains become non-superconducting. The percentage of non-superconducting grains increases with decreasing $\Gamma$. 
    
    (4) If the initial volume percentage of metallic grains is large enough, the system goes through a superconductor-metal percolation transition if the percentage of metallic grains with proximity effect destroyed is larger than the percolation threshold. If the initial percentage of metallic grains is not large enough, the system remains superconducting.
    
    (5) In the case when the system becomes a metal, a further metal-insulator transition occurs between metallic grains when $\Gamma$ decreases further. We note that as $\alpha_{MS}<<\alpha_{MM}$ the destruction of proximity effect should occur {\em before} the metal-insulator transition between the metallic grains takes place (see Fig.(1)). 
     The metal-insulator transition is also a percolation transition as the distribution of $\Gamma$ is random. 
     
    (6)Similarly, a superconductor-insulator transition occurs between superconducting grains when $\Gamma$ decreases further if the concentration of metallic grain is not high enough as described in (4).
    
    We note that the conductance of the system can become arbitrary large (small) when the system is close to the superconductor-metal (metal-insulator) percolation transition, in agreement with the large conductance variation observed experimentally when the anomalous metal state evolves between the superconductor and insulator\cite{rmp}. 
    
   \subsubsection{Discussion}
    In this paper we study a model of mixed superconductor and metallic grains regularly distributed on a square lattice with the nearest grains coupled by electron tunneling and show that the proximity effect through which superconductivity is induced on the metallic grains can be destroyed by quantum phase fluctuations if the charging energy $E_0$ is strong enough ($E_0>>J\frac{\Gamma}{\Delta_0} ln(\frac{\Delta_0}{\Gamma})$, see Eq.\ (\ref{ratio}))and the tunneling strengths $\Gamma$ is weak enough ($\Gamma<<\Delta_0$). Using the estimation $E_0\sim e^2/\epsilon L$ where $\epsilon$ is the dielectric constant and $L\sim$ size of grains\cite{gsize}, we see that proximity effect is destroyed when $L_c>>L$ where $L_c=L_0{\frac{\Delta_0}{\Gamma} ln(\frac{\Delta_0}{\Gamma}})$ where $L_0=e^2/J\epsilon\sim 10^{2-3}nm$\cite{gsize}.  Our model complements the model of isolated superconducting grains immersed in a metallic ocean\cite{fl,sz,sk} where superconductivity is suppressed by interaction only when distance between the grains is large enough.
    
     We caution that we have restricted ourselves to zero temperature and zero magnetic field in our paper and these effects have to be included in a a full microscopic theory for the anomalous metal state\cite{rmp, topo, pnas,univ,strunk}. We also note that experimentally charge-$2e$ carriers seem to be present in the anomalous metal state\cite{c2e} and forms the basis for alternative proposals of Bose-metal\cite{bm} or phase-glass\cite{pp} as the fundamental mechanism for the anomalous metal state. Dissipative charge-$2e$ carriers exist in the metal phase in our model as Cooper pairs in the superconductor grains remains intact even when proximity effect is destroyed. A detailed understanding of the role of charge-$2e$ carriers in our model and the differences between "homogeneous" and "granular" disordered superconductors is needed for a further understanding of the anomalous metal state.
     
    The role of metallic components in anomalous metal state can be tested by observing the transport properties of artificial two-dimensional (2D) structures composed of superconducting and metallic components\cite{art1,art2,art3}. We suggest to investigate 2D structures with superconductor and metal grains connected into alternating, parallel granular superconducting and metal strips. With strong enough charging energies, by adjusting the tunneling strengths between the underlying grains, the system will evolve from a uniform superconductor when tunneling is strong to an anisotropic system which is superconducting along the strips and metallic in the perpendicular direction when tunneling is weak and proximity effect is destroyed according to our analysis. 

  The author acknowledge special support from the School of Science, HKUST.



\subsection{Supplementary Materials}
\subsubsection{Electron Green's function}
  Using Eqs.\ (\ref{m}-\ \ref{gg}),  the electron Green's function $G_{S(M)k}$ is given in the self-consistent Born approximation by
  \begin{eqnarray}
  G_{Sk}^{-1}(i\omega) & = & \begin{pmatrix}
    i\omega+\xi_{k} & |\Delta_0| \\
   |\Delta_0|  & i\omega-\xi_{-k}
   \end{pmatrix}-4|t|^2\sum_{p}G_{Mp}(i\omega)  \\ \nonumber
   & = & \begin{pmatrix}
    i\omega+\xi_{k}-\Sigma_S(i\omega) & |\Delta_0|+D_S(i\omega) \\
   |\Delta_0|+D_S(i\omega)  & i\omega-\xi_{-k}-\Sigma_S(i\omega)
   \end{pmatrix}
  \end{eqnarray}
  and
  \begin{eqnarray}
  G_{Mk}^{-1}(i\omega) & = & \begin{pmatrix}
    i\omega+\xi_{k} & 0 \\
   0  & i\omega-\xi_{-k}
   \end{pmatrix}-4|t|^2\sum_{p}G_{Sp}(i\omega)  \\ \nonumber
   & = & \begin{pmatrix}
    i\omega+\xi_{k}-\Sigma_M(i\omega) & D_M(i\omega) \\
   D_M(i\omega)  & i\omega-\xi_{-k}-\Sigma_M(i\omega)
   \end{pmatrix}.
  \end{eqnarray}
  Therefore
  \begin{eqnarray}
  G_S(i\omega) & = & \frac{1}{V}\sum_k G_{Sk}(i\omega) \\ \nonumber
  & = & \frac{1}{V}\sum_k\frac{1} {\Omega_S^2-\xi_k^2-|\bar{D}_S|^2}\begin{pmatrix}
    \Omega_S-\xi_{k} & -\bar{D}_S \\
   -\bar{D}_S  & \Omega_S+\xi_{-k}
   \end{pmatrix}   \\ \nonumber
   & \sim & N_S(0)\int^D_{-D}\frac{d\xi} {\Omega_S^2-\xi^2-|\bar{D}_S|^2}\begin{pmatrix}
    \Omega_S-\xi & -\bar{D}_S \\
   -\bar{D}_S  & \Omega_S+\xi
   \end{pmatrix} 
  \end{eqnarray}
  where $\Omega_S=i\omega-\Sigma_S(i\omega)$, $\bar{D}_S=|\Delta_0|+D_S(i\omega)$
  and
  \begin{eqnarray}
  G_M(i\omega) & = &  \frac{1}{V}\sum_k G_{Mk}(i\omega)  \\ \nonumber
  & = & \frac{1}{V}\sum_k \frac{1} {\Omega_M^2-\xi_k^2-|\bar{D}_M|^2}\begin{pmatrix}
    \Omega_M-\xi_{k} & -\bar{D}_M \\
   -\bar{D}_M  & \Omega_M+\xi_{-k}
   \end{pmatrix}   \\ \nonumber
   & \sim & \int^D_{-D}\frac{d\xi  N_M(0)} {\Omega_M^2-\xi^2-|\bar{D}_M|^2}\begin{pmatrix}
    \Omega_M-\xi & -\bar{D}_M \\
   -\bar{D}_M  & \Omega_M+\xi
   \end{pmatrix} 
  \end{eqnarray}  
  where $\Omega_M=i\omega-\Sigma_M(i\omega)$, $\bar{D}_M=D_M(i\omega)$. The integrals can be evaluated directly to obtain Eq.\ (\ref{scg}).
  
    Writing $g_{S(M)}(i\omega\rightarrow\omega)=\omega\gamma_{S(M)}(\omega)$, it is straightforward to obtain Eq.\ (\ref{scg1}), with
    \begin{eqnarray}
    \label{c1}
    \Delta_M(\omega) & = & \frac{4|t|^2f_S(\omega)}{1-4|t|^2\gamma_S(\omega)}  \\ \nonumber
    \Delta_S(\omega) & = & \frac{\Delta_0+4|t|^2f_M(\omega)}{1-4|t|^2\gamma_M(\omega)}.
    \end{eqnarray}
    Eliminating $\gamma_{S(M)}(\omega)$ and $f_{S(M)}(\omega)$ using Eqs.\ (\ref{scg1}) and\ (\ref{c1}), we obtain the self-consistent equations
    \begin{subequations}
    \label{c2}
    \begin{equation}
        \label{c2a}
        \Delta_M(\omega)=\frac{i\Gamma_S\Delta_S(\omega)}{\sqrt{\omega^2-|\Delta_S(\omega)|^2}+i\Gamma_S}
    \end{equation}
    and
    \begin{equation}
    \label{c2b}
    \Delta_S(\omega)=\frac{\Delta_0}{1+\frac{i\Gamma_M}{\sqrt{\omega^2-|\Delta_M(\omega)|^2}}}+\frac{i\Gamma_S\Delta_M(\omega)}{\sqrt{\omega^2-|\Delta_M(\omega)|^2}+i\Gamma_M}.
    \end{equation}
    \end{subequations}
      It is easy to see from Eq.\ (\ref{c2a}) that $\Delta_M(\omega)\rightarrow\Delta_S(\omega)$ in the limit $\Gamma_M,\Gamma_S>>\Delta_0>\omega$ and the solution to Eq.\ (\ref{c2b}) is then $\Delta_S(\omega)=\Delta_0$.
      In the other limit $\Gamma_M,\Gamma_S<<\Delta_0 (>>\omega)$, the solutions given in Eq.\ (\ref{Ds}) in main text can be obtained similarly from Eq.\ (\ref{c2}). 
      
      For systems with identical superconductor grains, $\Delta_S(\omega)$ is determined by
      \[
      \Delta_S(\omega) = \frac{\Delta_0+4|t|^2f_S(\omega)}{1-4|t|^2\gamma_S(\omega)}.  \]
      It is straightforward to show that $\Delta_S(\omega)=\Delta_0$. For identical metallic grains $\Delta_0=0$ and 
      \[
    G_M(\omega)\sim-i\pi N_M(0)\begin{pmatrix}
    1  & 0  \\
    0 & 1
    \end{pmatrix},  \]
    i.e., $G=G_0$ and the Green's functions are not renormalized by electron tunneling for identical grains. 
      
     \subsubsection{Mean-field treatment of Action\ (\ref{action})}
      We consider the action\ (\ref{action}) on a regular array of grains as discussed in the main text. To evaluate the free energy we follow Refs.\cite{granular2, NL} and write
      \begin{equation}
        \label{compact}
    \phi_{i}(\tau)=\frac{2\pi n_i\tau}{\beta}+\phi_{pi}(\tau)
     \end{equation}
      where $\phi_{pi}(\tau)=\phi_p(\tau+\beta)$ is the periodic part of the phase field and $\frac{2\pi n_i\tau}{\beta}$ is the non-periodic part where $n_i$ is an arbitrary integer. To evaluate the free energy we have to evaluate the path integral over the periodic $\phi_{pi}(\tau+\beta)$ fields and summing over all possible values of integer $n_i$'s.
      
      A variational approach was used to evaluate the free energy associated with action\ (\ref{action}) by employing a trial action $S_{trial}=S(\{n_i\})+S(\{\phi_p\})$ where $S(\{n_i\})$ and $S(\{\phi_p\}$ are actions depending only on the integer fields $\{n_i\}$ and periodic fields $\{\phi_{pi}\}$, respectively\cite{granular2, NL}. $S(\{n_i\})$ and $S(\{\phi_p\}$ can be determined from the mean-field decomposition
      \begin{eqnarray}
      \label{mf}
      \langle(\frac{\partial}{\partial\tau}\Delta\phi_{ij}(\tau))^2\rangle & \rightarrow & \langle\frac{4\pi^2(\Delta n_{ij})^2}{\beta^2}\rangle_{S(n)} +\langle(\frac{\partial}{\partial\tau}(\Delta\phi_{pij}(\tau))^2\rangle_{S(\phi_p)}  \\ \nonumber
      \langle e^{i\Delta\phi_{ij}(\tau))}\rangle & \rightarrow & \langle e^\frac{2\pi i\Delta n_{ij}\tau}{\beta}\rangle_{S(n)}\langle e^{i\Delta\phi_{pij}(\tau))}\rangle_{S(\phi_p)}  \\ \nonumber
      & \sim & \langle\delta(\Delta n_{ij})\rangle_{S(n)}\langle e^{i\Delta\phi_{pij}(\tau))}\rangle_{S(\phi_p)}
      \end{eqnarray}
      etc., where $\Delta n_{ij}=n_i-n_j$, $\langle...\rangle_{S(n)/S(\phi_p)}$ denotes averages with respect to $S(\{n_i\})$ and $S(\{\phi_p\})$, respectively.
      
      The $\cos(\frac{\Delta\phi_{ij}(\tau)-\Delta\phi_{ij}(\tau')}{2})$ term is evaluated similarly with the further approximation\cite{granular2, NL}
      \begin{eqnarray}
      \label{sdiss}
      \langle e^{i\frac{\Delta\phi_{ij}(\tau)-\Delta\phi_{ij}(\tau')}{2}}\rangle & \sim & \langle e^{\frac{2\pi i\Delta n_{ij}(\tau-\tau')}{2\beta}}\rangle_{S(n)}\langle e^{i\frac{\Delta\phi_{pij}(\tau)-\Delta\phi_{pij}(\tau')}{2}}\rangle_{S(\phi_p)}  \\ \nonumber
      & \sim & \langle e^{\frac{2\pi i\Delta n_{ij}(\tau-\tau')}{2\beta}}\rangle_{S(n)}\times \\ \nonumber & & \langle(1-\frac{1}{2}(\frac{\Delta\phi_{pij}(\tau)-\Delta\phi_{pij}(\tau')}{2})^2)\rangle_{S(\phi_p)}.
      \end{eqnarray}
      
      With these approximations, we arrive at the mean-field trial actions\cite{NL}
      \begin{subequations}
      \begin{equation}
          \label{sn}
          S(\{n_i\})=\sum_{ij=i+\delta}\left(\frac{2\pi^2}{\beta E_0}(\Delta n_{ij})^2+\frac{\alpha}{2}|\Delta n_{ij}|-\beta J_{MF}\delta(\Delta n_{ij})\right)
      \end{equation}
      which is a modified classical Solid-On-Solid model with $\alpha=\alpha_{SM}$ for our mixed grain model and $\alpha=\alpha_{MM}$ for RSJ model and
      \begin{eqnarray}
          \label{sphi}
          S(\{\phi_p\} & = & \sum_{i,j=i+\delta}\left(\int_0^{\beta}d\tau\left(\frac{1}{2E_0}(\frac{\partial}{\partial\tau}\Delta \phi_{pij}(\tau))^2+\frac{J_{R}}{2}(\Delta \phi_{pij}(\tau))^2\right)\right.   \\ \nonumber
          & & +\frac{1}{8}\left.\int_0^{\beta}d\tau\int_0^{\beta}d\tau'\alpha(\tau-\tau')(\Delta\phi_{pij}(\tau)-\Delta \phi_{pij}(\tau'))^2\right),
      \end{eqnarray}
      where $\alpha(\tau)=\alpha(\frac{kT}{\sin(\pi kT\tau)})^2$.
      
     $J_{MF}$ and $J_R$ are mean-field parameters determined by minimizing the approximate free energy $F=F_0+{\beta}^{-1}\langle S-S_{trial}\rangle$, where $F_0$ is the free energy associated with $S_{trial}$ and the averages $\langle...\rangle$ are carried out with respect to the trial action $S_{trial}$.
      \end{subequations}
      We obtain the mean-field equations\cite{NL}
      \begin{eqnarray}
      \label{mfeq}
      J_{MF} & = & Je^{-\frac{1}{2}\langle|\Delta\phi_p|^2\rangle}  \\ \nonumber
      J_R & = & J_{MF}P_n(0)
      \end{eqnarray}
      where $J=J_{SM}$ for our mixed grain model and $J=J_{MM}$ for RSJ model. $P_n(m)=\langle\delta(|\Delta n_{ij}|-m)\rangle_{S(n)}$ is the probability that the nearest neighbor integer difference has magnitude $|n_i-n_j|=m$ in the integer action $S(\{n_i\})$,
      \begin{equation}
      \label{reno}
          \langle|\Delta\phi_p|^2\rangle=\frac{1}{2\beta}\sum_{i\omega_n}\frac{1}{\frac{\omega_n^2} {E_0}+J_R+\frac{\alpha}{4\pi}|\omega_n|}.
      \end{equation}
      
      Eqs.\ (\ref{mfeq}) and\ (\ref{reno}) with $P_n(0)=1$ was first obtained in Ref.\ (\cite{CK}) by considering only the periodic $\phi_{pi}(\tau)$ field. It was found that at zero temperature and for large enough $E_0$, the transition point where $J_R\rightarrow0$ depends only on the value of $\alpha$, with $\alpha_c\sim1$, corresponding to tunneling resistance $R=h/2e^2$, as illustrated in Fig.(1). The resulting system is in the insulator phase for periodic square lattice as $S(\{n_i\})$ is already in the rough phase when $J_R=0$\cite{NL}. For the mixed grain model we study, Eq.\ (\ref{reno}) should be replaced by
      \begin{equation}
      \label{reno1}
          \langle|\Delta\phi_p|^2\rangle=\frac{1}{2\beta}\sum_{i\omega_n, |\omega_n|>J_R}\frac{1}{\frac{\omega_n^2} {E_0}+J_R+\frac{\alpha}{4\pi}|\omega_n|}.
      \end{equation}
      as the expression is valid only in energy regime $\omega>\Delta_M$. As explained in the main text, $\Delta_M\rightarrow0$ when $J_R\rightarrow0$ and the cutoff in $\omega_n$ is unimportant as far as the superconductor-insulator phase transition is concerned. 
      

\begin{references}
\bibitem{t00} This can be seen most easily by replacing $t_{kp}^{(ij)}$ by $t_{kp}^{(ij)}\langle
e^{\frac{i}{2}\Delta\phi_{ij}(\tau)}\rangle$ in $T_0$ when we evaluate $G$.
\end{references}
\end{document}